\documentclass[%
 superscriptaddress,
preprint,
showpacs,preprintnumbers,
nofootinbib,
 amsmath,amssymb,
 aps,
 prd,
]{revtex4-1}

\pdfoutput=1

\usepackage{amssymb}
\usepackage{mathrsfs}
\usepackage{dsfont}
\usepackage{graphicx,subfigure}
\usepackage{enumitem}

\usepackage[colorlinks,
            linkcolor=black,
            anchorcolor=black,
            citecolor=black
            ]{hyperref}

\usepackage{xcolor}
\usepackage{multirow}

\newlength{\dinwidth}
\newlength{\dinmargin}
\newcommand{\GeV}{{\,\rm GeV}}
\newcommand{\diff}{{\rm d}}

\newcommand{\Br}{{\mathcal B}}

\begin{document}
\preprint{NCTS-PH/1801}

\title{Search for sterile neutrinos decaying into pions at the LHC}
\date{}

\author{Claudio O. Dib}
\email[E-mail: ]{claudio.dib@usm.cl}
\affiliation{Department of Physics and CCTVal, Universidad T\' ecnica Federico Santa Mar\'\i a, 
Valpara\'\i so  2340000, Chile}

\author{C.S. Kim}
\email[E-mail: ]{cskim@yonsei.ac.kr}
\affiliation{Department of Physics and IPAP, Yonsei University, Seoul 120-749, Korea}

\author{Nicol$\acute{\text{a}}$s A. Neill}
\email[E-mail: ]{nicolas.neill@gmail.com}
\affiliation{Department of Physics and CCTVal, Universidad T\' ecnica Federico Santa Mar\'\i a, 
Valpara\'\i so 2340000, Chile}

\author{Xing-Bo Yuan}
\email[E-mail: ]{xbyuan@cts.nthu.edu.tw}
\affiliation{Physics Division, National Center for Theoretical Sciences, Hsinchu 30013, Taiwan}

\date{\today}

\begin{abstract}
\noindent
We study the possibility to observe sterile neutrinos with masses in the range 5 GeV $< m_N < 20$ GeV at the LHC, using the exclusive semileptonic modes involving pions, namely $W\to \ell N \to n \pi \ell\ell$ ($n=1,2,3)$.
The two pion and three pion modes require extrapolations of form factors to large time-like $q^2$, which we do using vector dominance models as well as light front holographic QCD, with remarkable agreement. 
This mass region is difficult to explore with inclusive $\ell\ell jj$ modes or trilepton modes and impossible to explore in rare meson decays. While particle identification is a real challenge in these modes, vertex displacement due to the long living neutrino in the above mass range can greatly help reduce backgrounds. 
 Assuming a sample of $10^9$ $W$ bosons at the end of the LHC Run 2, these modes could discover a sterile neutrino in the above mass range or improve the current bounds on the heavy-to-light lepton mixings  by an order of magnitude, $|U_{\ell N}|^2\sim 2\times 10^{-6}$. Moreover, by studying the equal sign and opposite sign dileptons, the Majorana or Dirac character of the sterile neutrino may be revealed. 

\end{abstract}

\maketitle


\section{Introduction}

Neutrino oscillation data~\cite{Fukuda:1998mi,Ahmad:2002jz,Eguchi:2002dm} demonstrates that at least two active neutrinos are massive, 
which provides strong indication of new physics~\cite{GonzalezGarcia:2007ib,Mohapatra:2006gs,Barger:2003qi,Strumia:2006db} beyond 
the minimal Standard Model. In order to naturally explain the smallness of the observed neutrino masses, right-handed sterile neutrinos are usually introduced in the SM extensions, creating scenarios with a seesaw mechanism~\cite{Minkowski:1977sc,Yanagida:1979as,Ramond:1979py,GellMann:1980vs,Glashow:1979nm,Mohapatra:1979ia,Schechter:1981cv}. In the past decades, many experiments have searched for sterile neutrinos with masses in the broad range from eV to TeV ~\cite{Deppisch:2015qwa}.

Current searches for sterile neutrinos with mass $m_N$ above $100\GeV$~\cite{Chatrchyan:2012fla,Aad:2015xaa} are based on the inclusive processes 
$pp \to W^\ast X$, $W^\ast\to \ell^\pm  \ell^\pm jj$~\cite{Keung:1983uu, delAguila:2007qnc, Atre:2009rg, Das:2017gke}. For $m_N$ below $M_W$, the jets tend to fall below the necessary cuts that are imposed to reduce backgrounds, so these inclusive semileptonic modes cease to be useful. Also purely leptonic modes $W^{(\ast)}\to \ell\ell \ell\nu $ have been proposed~\cite{Cvetic:2012hd}. The purely leptonic modes~\cite{Dib:2015oka,Izaguirre:2015pga, Dib:2016wge,Dib:2017vux,Dib:2017iva,Dube:2017jgo} are also an alternative to 
$\ell^\pm  \ell^\pm jj$ at the LHC, as they are not affected by the hadronic backgrounds for $m_N$ below $M_W$, even though they have the problem of missing energy due to the undetectable final neutrino. Also, in these three-lepton modes one should not use modes with a pair of leptons with same flavor and opposite charge, such as $e^+ e^-$ or $\mu^+ \mu^-$, in order to avoid the background from radiative pair production. Nevertheless, for neutrino masses below $\sim 20$ GeV, the neutrino may live long enough to leave an observable displacement from its production to its decay point, a feature that will also help reduce such background~\cite{Helo:2013esa,Dib:2014iga,Cottin:2018kmq}. 

In this work we consider the exclusive semileptonic processes $W \to \ell N$, $N \to \ell \pi$, $\ell \pi \pi$ and $\ell \pi\pi\pi$, which are modes with no missing energy. Among these modes, we do not consider $N\to e^\pm\pi^\mp$ in the secondary process to avoid misidentification of electrons and pions. The most promising modes should be $W^+\to \mu^+ N$, $N\to \mu^+ n\pi $ ($n=1,2,\ldots)$ for a Majorana sterile neutrino, and $W^+\to e^+ N$, $N\to \mu^- n \pi$ for a Dirac neutrino. 

The exclusive single-pion mode $N\to\ell \pi$ is suppressed with respect to the inclusive process, modeled by the production of an open quark pair  $N \to \ell q \bar q^\prime$,  by a factor 
$f_\pi^2/m_N^2$, where $f_\pi$ is the pion decay constant. 
This suppression is clearly milder for lower masses of the sterile neutrino. Moreover, for lower $m_N$ the sterile neutrino width is also smaller, giving time for the neutrino to travel a measurable distance before decaying. This effect will show as a spatial vertex separation in the detector, a distinctive feature that helps reduce backgrounds~\cite{Helo:2013esa, Dib:2014iga}. Less suppressed than the decay into a single pion is the decay into two pions, i.e. $N\to \ell \pi \pi$, or even into three pions, because the two quarks from the weak vertex are not forced to be bound into a single pion wave function~\cite{Lepage:1979zb}. 
Although there is a suppression from the 3-particle or 4-particle phase space, it is not as strong as the wave function restriction of the single pion case. 

As stated above, at the LHC the inclusive mode $\ell^\pm\ell^\pm jj$ is sensitive to neutrino masses $m_N$ above $100\GeV$ and the purely leptonic $\ell^\pm \ell^\pm\ell^{\prime \mp}\nu$ modes cover the region below $M_W$ down to about $20\GeV$. Complementary, for $m_N$  below 5 GeV, rare decays of  $B$, $D$ and $K$ mesons or tau leptons are more 
appropriate~\cite{Cvetic:2010rw,Aaij:2012zr,Cvetic:2017vwl,Dib:2000wm,Kim:2017pra}.  In this work, we focus on neutrino masses within the window from 5 GeV to 20 GeV,  using the exclusive processes $W^+ \to \ell^+ N$ followed by the pionic decays $N\to \pi^- \ell^+$, $N \to \pi^0 \pi^- \ell^+$ and $N \to \pi^-\pi^- \pi^+ \ell^+$. These processes can give the LHC the ability to explore the corresponding mass region \footnote{One caveat, however, is that the identification of pions and their separation from Kaons may be a formidable challenge for current experiments.}.

This article is organized as follows. In section.~\ref{sec:decay}, we calculate the pionic decays of a sterile neutrino. In section.~\ref{sec:numerical}, we detail our  numerical results and discussions. We conclude in section.~\ref{sec:conclusion}.

\section{Pionic decays of a sterile neutrino}\label{sec:decay}

The leptonic sector in a generic SM extension includes one 
or more extra neutral lepton singlets, $N$,
in addition to the three generations of left-handed SM $SU(2)_L$ lepton doublets.
  The neutral lepton singlets $N$ are known as \emph{sterile} neutrinos, because they do not directly interact with other SM particles in the absence of any mixing with the active neutrino sector. After  electroweak symmetry breaking, the gauge interaction of the sterile neutrino with the SM fields can be written as
\begin{align}\label{eq:Lagrangian}
  \Delta\mathcal L=-\frac{g}{\sqrt{2}} U_{\ell N}^* W_\mu^+ \bar N \gamma^\mu P_L \ell 
  + h.c.,
\end{align}
where $g$ denotes the weak coupling constant and $U_{\ell N}$ the mixing matrix between the active and sterile neutrinos. 
Here the sterile neutrino can be a Dirac or Majorana fermion.

At the LHC, sterile neutrinos with masses around $5\sim 20\GeV$ will be mainly produced from the decay of on-shell $W$ bosons. From the lagrangian in Eq.~\eqref{eq:Lagrangian}, the decay rate $W^+\to \ell^+ N$ can be easily calculated. Neglecting the lepton mass, the branching ratio ${\cal B}(W^+\to \ell^+ N)$ is:
\begin{align}
  {\cal B}(W^+ \to \ell^+ N) \equiv \frac{\Gamma(W^+\to \ell^+ N)}{\Gamma_W} =  \frac{G_F}{\sqrt{2}} \frac{M_W^3}{12\pi \Gamma_W} |U_{\ell N}|^2 \left( 2+\frac{m_N^2}{M_W^2}\right)\left( 1- \frac{m_N^2}{M_W^2}\right)^2,
\end{align}
where $\Gamma_W \simeq 2.085$ GeV is the total decay width of the $W$ boson~\cite{Patrignani:2016xqp}.
From here, the heavy neutrino $N$ can decay in several modes, depending on its mass. Here we are interested in the decays into pions, namely $N \to \pi^\mp \ell^\pm$, $N \to \pi^0 \pi^\mp \ell^\pm$  and $N \to \pi^\mp \pi^\mp \pi^\pm \ell^\pm$. Both charged modes will occur for a Majorana $N$, while for a Dirac $N$ 
only the $N$ decays into a negative charged lepton will be produced.

\subsection{The decay $\boldsymbol{N \to \pi^- \ell^+}$:}

The two-body decays of a heavy neutrino have been studied in detail~\cite{Atre:2009rg}. 
The mode  $N \to \pi^- \ell^+$ is a charged current
process:
%
\begin{align}\label{eq:width:1pi}
\Gamma(N \to \pi^- \ell^+)=&\frac{G_F^2}{16\pi}f_\pi^2 \lvert V_{ud} \rvert^2 \lvert U_{\ell N} \rvert ^2\,  m_N^3 \lambda^{1/2} (1,m_\ell^2/m_N^2, m_{\pi^-}^2/m_N^2) \\
&\qquad\quad
\times 
 \left[\left (1+ \frac{m_\ell^2}{m_N^2}-\frac{m_{\pi^-}^2}{m_N^2}\right) \left( 1+\frac{m_\ell^2}{m_N^2}\right) - 4\frac{m_\ell^2}{m_N^2}\right] .
 \nonumber
\end{align}
For an unpolarized $N$, the decay into the opposite charge mode $N\to \pi^+\ell^-$ is the same as above.
In an obvious notation, $m_{\pi^-}$ and $m_N$ denote the mass of the charged pion and sterile neutrino, respectively; $V_{ud}$ is the CKM matrix and $f_\pi$ is the pion decay constant; the function $\lambda (x,y,z)$ is defined as $\lambda(x,y,z)=x^2+y^2+z^2-2(xy+yz+zx)$. Now, the formation of a single pion in the final state is relatively suppressed with respect to multi pion modes, because it requires the two produced quarks to remain close together. Indeed, we can compare the above expression with that of open quark production, which is an estimate of the inclusive rate. Neglecting the masses in the final state, the rate is $
\Gamma(N \to \ell \bar q q^\prime) \sim (G_F^2/64 \pi^3) |V_{q q'}|^2|U_{\ell N}|^2 m_N^5$, 
and therefore the suppression factor of the exclusive $\ell \pi$ mode is approximately:

\begin{equation}
R \sim  \frac{\Gamma(N \to \ell^-  \pi^+)}{\Gamma(N \to \ell^-  u\bar d)} \sim 4\pi^2 \frac{f_\pi^2}{m_N^2}
\label{singlepion}
\end{equation}
For $m_N=10$ GeV, this suppression is $R\sim 0.6\%$. Consequently, in order to have a larger rate, we should also consider the decays with two and three pions in the final state.

\subsection{The decay $\boldsymbol{N \to \pi^0 \pi^- \ell^+}$:}\label{sec2b}

The decay into two pions, $N\to \pi^0 \pi^- \ell^+$, is similar to the $tau$ lepton decay $\tau^- \to \pi^0 \pi^- \nu_\tau$ in terms of their interaction lagrangian and Feynman diagram, except for the lepton flavor and charge. Indeed, the hadronic current in both decays has the same expression in terms of form factors. However, one must be aware that the kinematic range for the form factor in the $N$ decays is extended to higher $q^2$, so an extrapolation of the $tau$ form factor will be required.  
Considering the above, the differential decay rate for $N\to \pi^0 \pi^-\ell^+$ can be written as
\begin{align}\label{eq:width}
  \frac{\diff\Gamma(N \to \pi^0 \pi^- \ell^+)}{\diff s}=&\frac{\Gamma_N^0 \lvert V_{ud} \rvert^2 \lvert U_{\ell N} \rvert^2 }{2m_N^2} \frac{3s^3\beta_\ell \beta_\pi}{2m_N^6}F_-(s)^2 \\
&
\times 
\left[ \beta_\ell^2 \left( \frac{(\Delta m_\pi^2)^2}{s^2} - \frac{\beta_\pi^2}{3}\right) + \left( \frac{(m_N^2 - m_\ell^2)^2}{s^2}-1\right) \left( \frac{(\Delta m_\pi^2)^2}{s^2} + \beta_\pi^2 \right) \right], \nonumber
\end{align}
where   $\Gamma_N^0 =G_F^2m_N^5/(192\pi^3)$, $s = (p_{\pi^0}+p_{\pi^+})^2$,
$\Delta m_\pi^2 \equiv m_{\pi^+}^2 - m_{\pi^0}^2$, $\beta_\ell =\lambda^{1/2}(1,m_\ell^2/s,m_N^2/s)$, 
$\beta_\pi =\lambda^{1/2}(1,m_{\pi^+}^2/s,m_{\pi^0}^2/s)$, and 
$F_-(s)$ is the hadronic form factor of the charged current, defined by
\begin{align}
  \langle \pi^-(p) \pi^0(p^\prime) \vert \bar d \gamma_\mu u \vert 0 \rangle &= \sqrt{2} F_-(s) (p-p^\prime)_\mu  .
\label{eq:FF:-}
\end{align}  

The decay rate is then obtained after integrating over $s$: 
\begin{align}
  \Gamma(N \to \pi^0 \pi^- \ell^+)   = \int_{s_-}^{s_+} \diff s \ \frac{\diff\Gamma (N \to \pi^0 \pi^- \ell^+)}{ \diff s } ,
\end{align}
within the limits $s_- = (m_{\pi^-} + m_{\pi^0})^2$ and $s_+ = (m_N - m_\ell)^2$. 
It is easy to check that, after the replacement $(N,\ell^+)\to (\tau^-,\nu_\tau)$, our expression coincides with that of the decay width for $\tau^- \to \pi^0 \pi^- \nu_\tau$ in \cite{Cirigliano:2001er,Cirigliano:2002pv}. The main challenge here is to estimate the form factor $F_-(s)$. For low $s$, the form factor can been calculated in the framework of Chiral Perturbation Theory (ChPT)~\cite{Guerrero:1997ku}. 
In the time-like region, i.e.\  $s>0$, $F_-(s)$ is experimentally known from $\tau^- \to \pi^- \pi^0 \nu_\tau$ decay~\cite{Fujikawa:2008ma}, but only in the limited range $2m_\pi<\sqrt{s}<m_\tau$.
On the other hand, the neutral form factor $F_0(s)$, defined by
\begin{align}    \langle \pi^-(p) \pi^+(p^\prime) \vert \bar u \gamma_\mu u \vert 0 \rangle= - \langle \pi^-(p) \pi^+(p^\prime) \vert \bar d \gamma_\mu d \vert 0 \rangle &= F_0(s) (p-p^\prime)_\mu , \label{eq:FF:0}
\end{align}
which is related to $F_-(s)$ by conservation of the vector $SU(2)$ isospin current (CVC) as $F_0(s) = F_-(s)$, 
can be obtained from current $e^+ e^- \to \pi^+ \pi^-$ data~\cite{Lees:2012cj}  up to  $\sqrt{s}<3\GeV$. In our case, we need to know the form factor up to $m_N \sim 20$ GeV, so we will need to do an extrapolation. 
For this purpose we consider two theoretical models: one is the vector meson dominance model (VDM) used by the BaBar collaboration \cite{Lees:2012cj}, and another is based on light front holographic 
QCD (LFH) \cite{Brodsky:2014yha}. 
The VDM parametrization for $F_0(s)$, which includes $\rho$-$\omega$ mixing, is given by\,\cite{Lees:2012cj}
\begin{align}\label{eq:parameterization}
  F_0(s)=&\frac{1}{1+c_{\rho^\prime} + c_{\rho^{\prime\prime}} + c_{\rho^{\prime\prime\prime}} } 
  \biggl( 
  {\rm BW}_\rho^{\rm GS}(s,m_\rho, \Gamma_\rho)  \frac{1+c_\omega {\rm BW}_\omega^{\rm KS}(s,m_\omega,\Gamma_\omega)   }{1+c_\omega} 
 \\ &+ \ 
   c_{\rho^\prime} {\rm BW}_{\rho^\prime} ^{\rm GS}(s,m_{\rho^\prime}, \Gamma_{\rho^\prime}) 
   + 
   c_{\rho^{\prime\prime}} {\rm BW}_{\rho^{\prime\prime}} ^{\rm GS}(s,m_{\rho^{\prime\prime}}, \Gamma_{\rho^{\prime\prime}})    
   + 
   c_{\rho^{\prime\prime\prime}} {\rm BW}_{\rho^{\prime\prime\prime}} ^{\rm GS}(s,m_{\rho^{\prime\prime\prime}}, \Gamma_{\rho^{\prime\prime\prime}})  
   \nonumber
   \biggr),
\end{align}
where the Breit-Wigner functions ${\rm BW}_i (s, m_i, \Gamma_i )$ and numerical values of all the coefficients   fitted to the $e^+ e^- \to \pi^+ \pi^-$ data in the region $\sqrt{s} <3\GeV $, can be found in \cite{Lees:2012cj}
(see also~\cite{Kang:2013jaa}). 
Here no vector mesons heavier than $3\GeV$ contribute to $F_0(s)$, because  no light unflavored vector mesons heavier than $3\GeV$ have been observed yet, $q \bar c$ and $q \bar b$ mesons  (e.g. $D^*(2007)$ and $B^*(5324)$) do not decay to $\pi \pi$ through strong interactions and are therefore suppressed, and the decays of $c \bar c$ and $b \bar b$ mesons (e.g. $J/\psi$ and $\Upsilon$) into  two pions are OZI suppressed.

Since isospin is broken by the small quark mass difference $m_d-m_u$, the mesons $\rho^0$ ($I=1$) and $\omega$ ($I=0$) are not exact isospin eigenstates, but they mix with each other \cite{O'Connell:1995wf}. This feature results in a small structure near $\sqrt{s}=0.78\GeV$ in the $e^+ e^- \to \pi^+ \pi^-$ spectrum. For the form factor $F_0(s)$, this effect is included in Eq.\,\eqref{eq:parameterization}. However, since there is no charged $\omega$ meson, no $\rho-\omega$ mixing should  be present in $F_-(s)$ and, moreover, since we are obtaining $F_-(s)$ from the CVC isospin relation $F_-(s)=F_0(s)$, isospin violation effects should be omitted. Nevertheless,
since the parameters in $F_0(s)$ were found by fitting the data altogether and this isospin violation effect is very small, we kept the $\omega$ term in  $F_-(s)$.

Concerning the holographic QCD formalism \cite{Brodsky:2014yha}, the form factor is dominated by the twist-2 and twist-4 terms (in this formalism, the twist `$\tau$' corresponds to the number of partons in the light-front Fock state and is also related to the number of poles in the form factor, which is $\tau-2$). In this approximation, the form factor is:
\begin{align}
F_\pi (s) = (1-\gamma) \ F_{\tau=2}(s) + \gamma \ F_{\tau=4} (s),
\end{align}
where $\gamma = 12.5\,\%$ is the probability of the twist-4 component. The two components are: 
\begin{align}
F_{\tau=2}(s) = \frac{m_\rho^2}{(m_\rho^2 - s - i \sqrt{s}\ \Gamma_\rho)}
\end{align}
and
\begin{align}
F_{\tau=4}(s) = \frac{m_\rho^2 m_{\rho_1}^2 m_{\rho_2}^2}
{(m_\rho^2 - s - i \sqrt{s}\ \Gamma_\rho)
(m_{\rho_1}^2 - s - i \sqrt{s}\ \Gamma_{\rho_1})
(m_{\rho_2}^2 - s - i \sqrt{s}\ \Gamma_{\rho_2})}, 
\end{align}
and the following parameters were used for their best fit up to $s\sim 3$ GeV$^2$: $m_\rho = 0.775\ \textrm{GeV}$, $\Gamma_\rho = 149\ \textrm{MeV}$, 
$m_{\rho_1} = 1.343 \ \textrm{GeV}$, $ \Gamma_{\rho_1} = 360 \ \textrm{MeV}$, $m_{\rho_2} = 1.733\ \textrm{GeV}$, $ \Gamma_{\rho_2} = 160\ \textrm{MeV}$. 
According to Ref.\ \cite{Brodsky:2014yha}, their fit agrees with data up to $s\sim 6$ GeV$^2$.

\begin{figure}[h]
  \centering
  \subfigure{\label{fig:1}\includegraphics[width=0.45\textwidth]{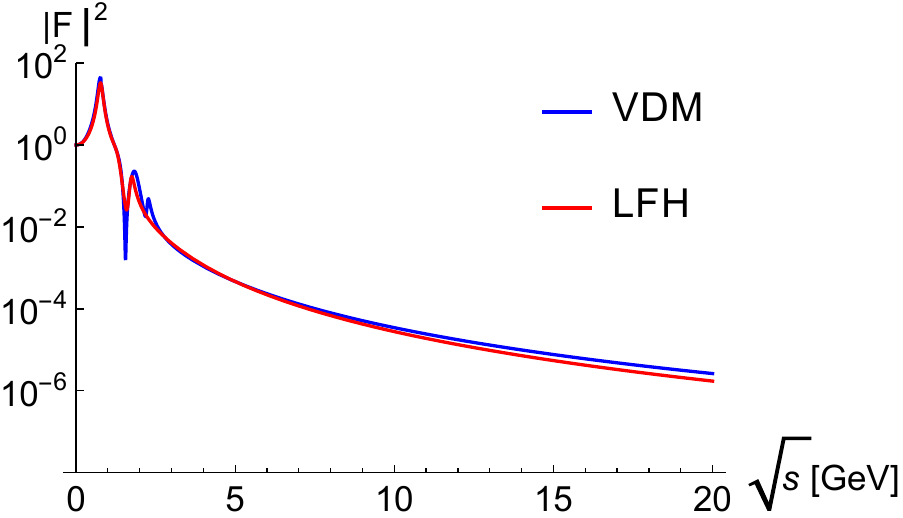}}
  \quad
  \subfigure{\label{fig:2}\includegraphics[width=0.45\textwidth]{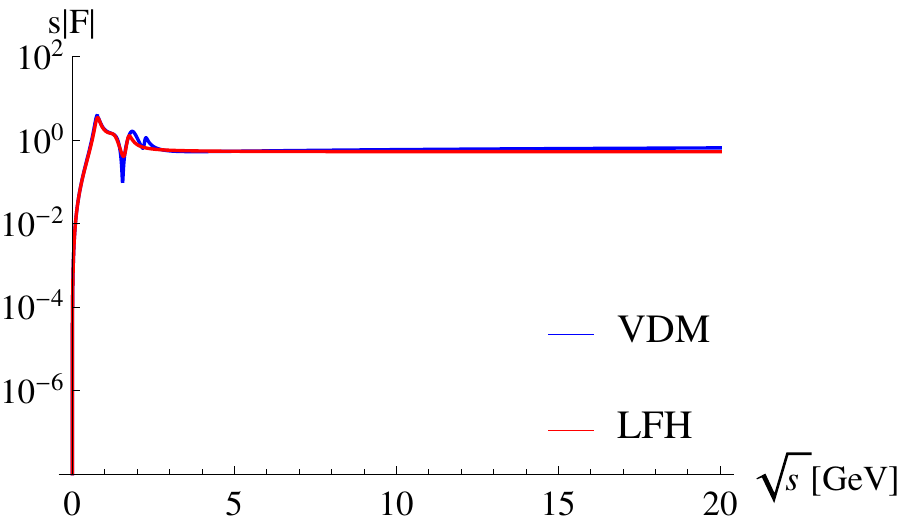}}
  \caption{\baselineskip 3.0ex
The pion form factor according to the vector-dominance model (blue) and  Light-front holography (red), extrapolated  up to $\sqrt{s} < 20$ GeV. The right figure shows the asymptotic behavior $s F(s) \to constant$ at large $s$.}
  \label{plot2FF}
\end{figure}

The pion form factor according to these two models  is shown in Fig.\  \ref{plot2FF}, including the extrapolated region up to $\sqrt{s} = 20$ GeV.
Although the VDM model gives a more detailed structure at low $\sqrt{s}$ than the LFH formalism, the overall coincidence is outstanding. Concerning the extrapolation to larger $\sqrt{s}$,  e.g.\ at $\sqrt{s}= 15$ GeV, which is an extrapolation far beyond the resonant region, we obtain $F_\pi^{\rm (VDM)} = 2.8\times 10^{-3}$ and $F_\pi^{\rm (LGH)} = 2.3 \times 10^{-3}$, a discrepancy no more than 20\%, quite good an agreement considering it is far into the tails  of the wave functions. Also, in both models one finds the QCD behavior $s F(s)\to constant$ at large $s$, as shown in Fig.\ \ref{plot2FF} (right). Since our purpose is to estimate 
$N\to  \pi\pi \ell$ for $m_N$ within the range 5 GeV to 
20~GeV, it is necessary to have a good estimate of the form factor extrapolated beyond current data. Fortunately the  extrapolated region (large $s$) is where theory is more reliable, and the most difficult part to address theoretically  is the resonance region where data is available. Indeed, as the above study shows,  the extrapolation with two different treatments give quite similar results, so we expect this estimate to be reliable for the purpose of our work.

\subsection{The decay $\boldsymbol{N \to \pi^- \pi^- \pi^+ \ell^+}$:}

In much the same way as in the two-pion mode of Section II.B, the  differential decay rate of the general hadronic decay $N \to h_1 h_2 h_3 \ell^+$ can be written in terms of form factors with an expression identical to that of the tau decay $\tau^- \to h_1 h_2 h_3 \nu_\tau$ \cite{Dumm:2009kj}, again provided that the form factors are extrapolated to larger values of $q^2$.  
Let us first denote the momentum and mass of the hadron $h_i$ ($i=1,2,3$) by $p_i$ and $m_i$ respectively, and define the momentum of the hadronic part
$q^\mu = (p_1+p_2+p_3)^\mu$. 
The differential decay rate is then:

\begin{align}\label{eq:width:3h}
  \frac{\diff \Gamma(N \to h_1 h_2 h_3 \ell^+)}{\diff q^2} =& \frac{G_F^2|V_{ud}|^2|U_{\ell N}|^2}{128(2\pi)^5\ m_N^3}  \lambda^{1/2}(1, m_N^2/q^2 ,m_\ell^2/q^2)  
    \bigg[ \left( \frac{(m_N^2-m_\ell^2)^2}{q^2} - m_N^2 - m_\ell^2 \right) \omega_{SA} (q^2)
  \nonumber \\
  &
    + \frac{1}{3}\left( \frac{(m_N^2 - m_\ell^2)^2}{q^2}+ m_N^2 + m_\ell^2  -2q^2\right)(\omega_A(q^2) + \omega_B(q^2)) \bigg] . 
\end{align}

%

Here the functions $\omega_i(q^2)$, $i=${\footnotesize\it  A, B, SA}, are defined as~\cite{Kuhn:1992nz}:
\begin{equation}
\omega_i(q^2) =\int_{s_3 ^-}^{s_3^+} \diff s_3 \int_{s_2^-}^{s_2^+ } \diff s_2 \, W_i(q^2,s_2, s_3),
\end{equation}
with the variables $s_3= (p_1+p_2)^2$ and $ s_2=(p_1 +p_3)^2$, and the integration limits: 
\begin{align}
  s_{2}^{ \pm}=& \frac{1}{4s}\left[ (q^2+m_1^2-m_2^2-m_3^2)^2 - (\lambda^{1/2}(q^2,s,m_3^2)\mp \lambda^{1/2}(m_1^2,m_2^2,s))^2 \right],
\\
s_{3}^{ -}=& (m_1+m_2)^2, \quad 
s_{3}^{ +}=(\sqrt{q^2}-m_3)^2 .
\end{align}
The integrands $W_i(q^2,s_2,s_3)$ are the following Lorentz invariants:
\begin{align}
 W_A(q^2,s_2,s_3) &=-\Big|V_1^\mu \ F_1^A(q^2,s_2,s_3) + V_2^\mu \  F_2^A(q^2,s_2,s_3)\Big|^2, \nonumber\\
 W_B(q^2,s_2,s_3)&=-\Big|V_3^\mu\  F_3^V(q^2,s_2,s_3)\Big|^2,\nonumber\\
  W_{SA}(q^2,s_2,s_3)&=+\Big|q^\mu \ F_4^P(q^2,s_2,s_3)\Big|^2 .
\end{align}

Here $V_1^\mu$, $V_2^\mu$ and $V_3^\mu$ are the kinematical vectors:
\begin{align}
  V_1^\mu =\left (g^{\mu\nu}-\frac{q^\mu q^\nu}{q^2}\right)  (p_1-p_3)_\nu ,
  \quad 
  V_2^\mu =\left (g^{\mu\nu}-\frac{q^\mu q^\nu}{q^2}\right)  (p_2-p_3)_\nu ,
  \quad
  V_3^\mu = i \varepsilon^{\mu \alpha \beta \gamma} p_{1\alpha} p_{2\beta} p_{3\gamma}  , 
 \end{align} 
while $F_1^A(q^2, s_2,s_3)$, $F_2^A(q^2, s_2,s_3)$, $F_3^V(q^2, s_2,s_3)$ and $F_4^P(q^2, s_2,s_3)$, are the form factors that appear in  the hadronic current matrix element for the decay  
$N \to h_1 h_2 h_3 \ell^+$ \cite{Dumm:2009kj}:
\begin{align}\label{eq:FF:3h}
  \langle h_1(p_1) h_2(p_2) h_3(p_3) | (V-A)^\mu | 0 \rangle = \ 
  & V_1^\mu \ F_1^A (q^2, s_2,s_3)  + V_2^\mu  \ F_2^A (q^2, s_2,s_3)
  \nonumber\\ 
&  + V_3^\mu  \ F_3^V (q^2, s_2,s_3)  + q^\mu  \ F_4^P (q^2, s_2,s_3).
\end{align}

 In this parametrization, the form factors $F_1^A$ and $F_2^A$ correspond to axial vector transitions ($J^P = 1^+$), while 
 $F_3^V$  and $F_4^P$ to vector  ($J^P = 1^-$) and pseudoscalar ($J^P=0^-$) transitions, respectively.
  
\begin{figure}[t]
  \centering
  \includegraphics[width=0.45\textwidth]{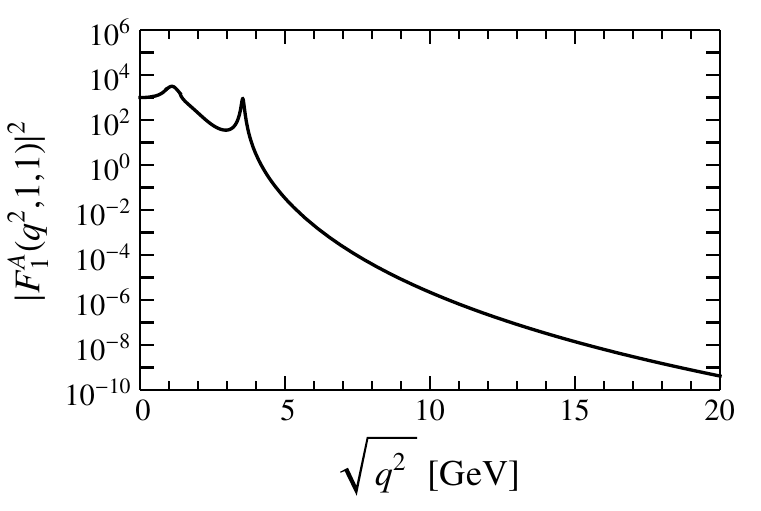}
  \includegraphics[width=0.45\textwidth]{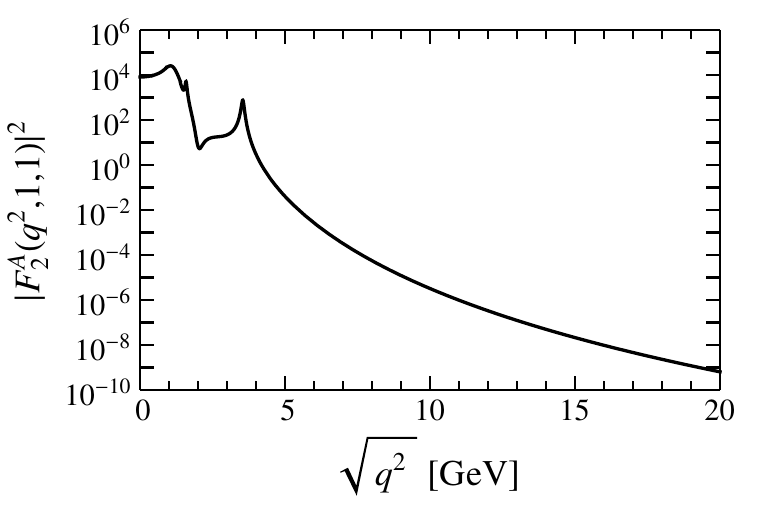}
  \caption{Form factor $F_1^A(q^2,s_2,s_3)$ and $F_2^A(q^2,s_2,s_3)$ at $s_2=s_3=1 \GeV^2$. The form factors in the region $q^2<m_\tau^2$ are obtained from the CLEO $\tau$ decay data, and extrapolated to the $q^2 > m_\tau^2$ region.}
  \label{fig:FF:3pi}
\end{figure}

Now, in the $N \to \pi^- \pi^- \pi^+ \ell^+$ decay, the partial conservation of the axial current (PCAC) implies that the pseudoscalar form factor $F_4^P$ is proportional to $m_\pi^2/q^2$, and therefore suppressed in most of the kinematical region\,\cite{Decker:1993ay,GomezDumm:2003ku,Dumm:2009va}. Moreover, in the isospin symmetry limit, $G$-parity conservation leads to $F_3^V=0$~\cite{Kuhn:1992nz}. Therefore, we only need to consider the two axial form factors $F_1^A$ and $F_2^A$. For  $q^2<m_\tau^2$ these two form factors were first obtained from the CLEO $\tau$ decay data, and then can be extrapolated to the $q^2>m_\tau^2$ region by using an appropriate parametrization. There are many parametrizations of the form factors in the literature. We chose the so-called CLEO parametrization\,\cite{Asner:1999kj}, which is used in the \texttt{TAUOLA} package \,\cite{Was:2015laa, Nugent:2013hxa}. This parametrization includes the transitions between various intermediate resonances $a_1(1260)/a_1^\prime(1640) \to \pi + f_0(500), f_2(1270), f_0(1370), \rho(770),$ and 
$\rho^\prime(1450)$. 
Just as in the $N \to \pi^0 \pi^- \ell^+$ decay, meson resonances heavier than the $\tau$ lepton do not contribute to the form factors. At $s_2=s_3=1\GeV^2$, the numerical results of $F_1^A(q^2, s_2,s_3)$ and $F_2^A(q^2, s_2,s_3)$ are shown in Fig.~\ref{fig:FF:3pi}. 

The decay rate is obtained after integration over $q^2$:
\begin{align}
  \Gamma(N \to h_1 h_2 h_3 \ell^+) = \int_{q_-^2}^{q_+^2} dq^2 \, \frac{\diff \Gamma(N \to h_1 h_2 h_3 \ell^+)}{\diff q^2},
\end{align}
where the integration limits are:
\begin{align}
 q_-^2=(m_1+m_2+m_3)^2, 
 \quad
 q_+^2 =(m_N-m_\ell)^2. 
\end{align}
 One can check that, after the replacement $(N,\ell^+) \to (\tau^-,\nu_\tau)$, our formulae coincide with those for the tau lepton decay  rate $\tau^- \to h_1 h_2 h_3 \nu_\tau$ given in \cite{Dumm:2009kj}.

\section{Numerical analysis}\label{sec:numerical}

With the theoretical framework described in the previous section, we want to estimate the exclusive semileptonic decay rates of $N$ 
into $\pi \ell$, $2\pi \ell$ and $3\pi\ell$, for a neutrino $N$ of mass in the range 5 to 20 GeV, produced at the LHC in the process
$W \to \ell N$. These exclusive modes could be observed in $pp$ collisions provided the pions can be identified and the  background can be reduced considering the spatial displacement between the production and decay vertices of $N$. This displacement could be observable for $m_N$ below 
20 GeV~\cite{Helo:2013esa,Dib:2014iga}.   
We first study the invariant mass distributions for the two-pion and three-pion modes, for two benchmark neutrino masses, $m_N=$ 5 GeV and 15 GeV, and then study the integrated rates of the single pion, two-pion and three-pion modes as a function of $m_N$. 

\begin{figure}[h]
  \centering
  \subfigure{\label{fig:1}\includegraphics[width=0.45\textwidth]{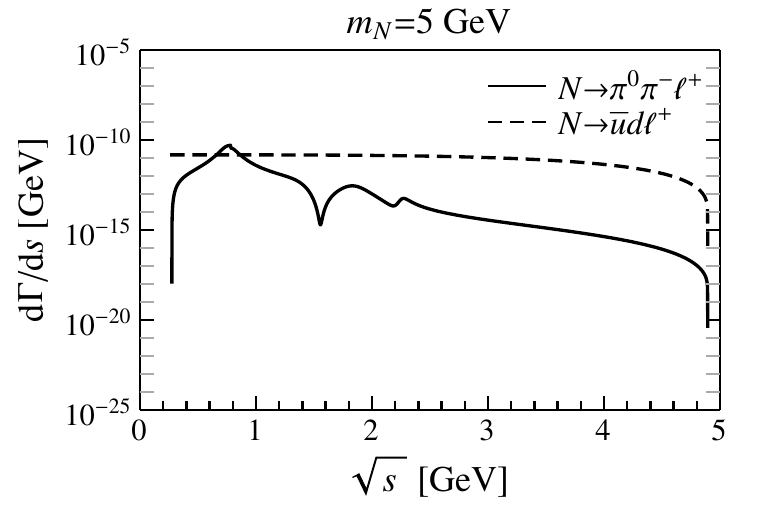}}
  \quad
  \subfigure{\label{fig:2}\includegraphics[width=0.45\textwidth]{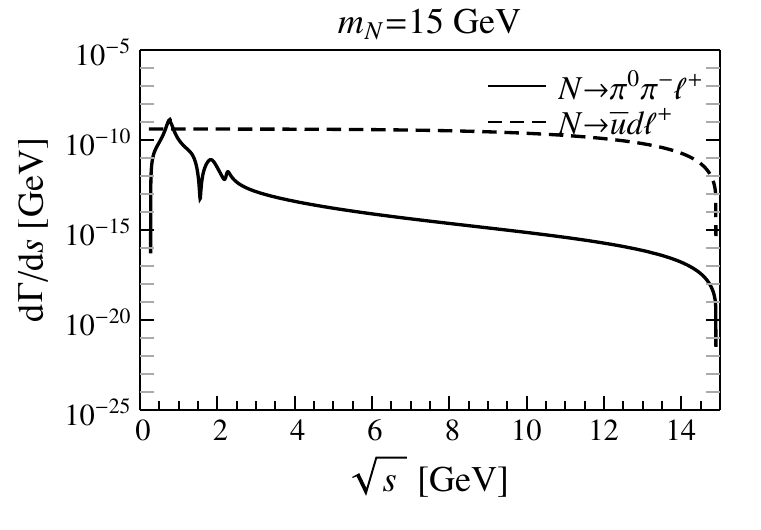}}
  \caption{\baselineskip 3.0ex
Differential decay width for the $N \to \pi^0 \pi^- \ell^+$ and $N \to \bar u d \ell^+$ decay with $m_N=5\GeV$ (left) and $m_N=15\GeV$ (right), where $\sqrt s$ denotes the invariant mass $M(\pi^0\pi^-)$ and $M(\bar u d)$ respectively. $|U_{\ell N}|=1$ is assumed.}
  \label{fig:width:diff}
\end{figure}

In Fig.\,\ref{fig:width:diff} we show the $M(\pi^0 \pi^-)$ distribution for the $N \to \pi^0 \pi^- \ell^+$ decay, and compare it with the $M(\bar u d)$ distribution for the open quark decay $N \to \bar u d \ell^+$, for the cases of $m_N=5\GeV$ and $m_N=15\GeV$. In the case of $m_N=5\GeV$ (left figure), the shape of the distribution is dominated by the $\rho$, $\rho^\prime$, $\rho^{\prime\prime}$, and $\rho^{\prime\prime\prime}$ resonances. The dip near $1.6\GeV$ can be explained by the interference between the $\rho^\prime$ and $\rho^{\prime\prime}$ contributions. In the case of $m_N=15\GeV$ (right figure), the distribution is similar to the former case.  
As explained in Section II.B, the form factor in the $\sqrt s > 3\GeV$ region is obtained from a na\"\i ve extrapolation, which results in the shape shown.

Within the quark-hadron duality~\cite{Shifman:2000jv}, the open quark process $N \to \bar u d \ell^+$ is considered to be the same as the inclusive semileptonic process. 
As shown in Fig.~\ref{fig:width:diff}, the exclusive mode is much smaller than the inclusive process in the invariant mass region far from the $\rho$ meson peak, while they are of comparable magnitude in the region near the peak. Moreover, considering the integration over $\sqrt{s}$ in these figures, one can notice that  the total decay width of $N \to \pi^0 \pi^- \ell^+$ is predicted, as it should, to be smaller than the inclusive estimate $N \to \bar u d \ell^+$, and this difference is enhanced for larger $m_N$, as there will be more modes with larger hadron multiplicities as $N$ is heavier.

\begin{figure}[h]
  \centering
  \includegraphics[width=0.49\textwidth]{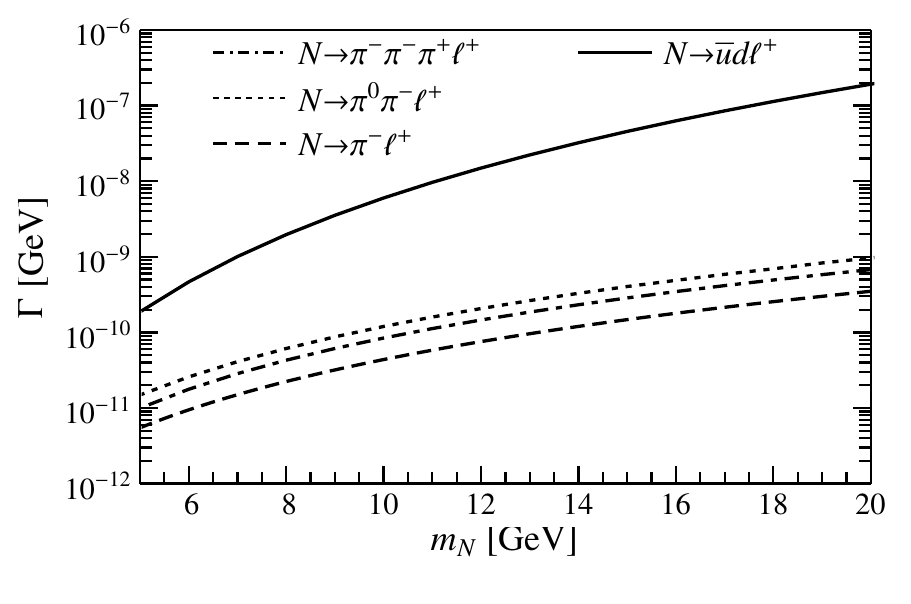} 
  \includegraphics[width=0.49\textwidth]{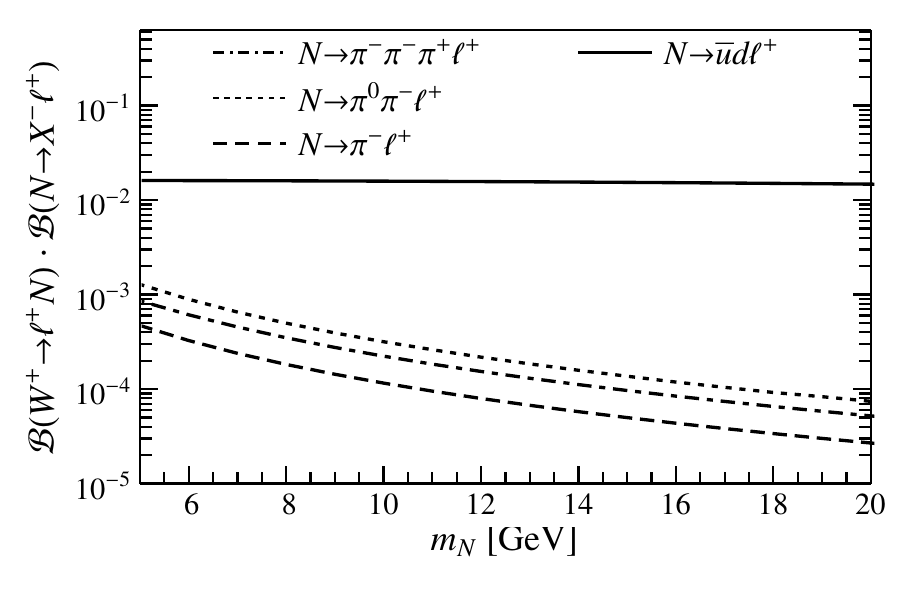}
  \caption{\baselineskip 3.0ex
 Canonical decay rates for $N \to \pi^- \ell^+$, $N \to \pi^0 \pi^- \ell^+$, $N \to \pi^- \pi^- \pi^+ \ell^+$ and $N \to \bar u d \ell^+$ (left) and canonical branching ratios for the full processes $W^+\to \ell^+\ell^+ n \pi$ (right) as a function of the neutrino mass $m_N$, with all mixing factors $|U_{\ell N}|$ removed. To obtain the actual values, the canonical values must be multiplied by the factor 
 $|U_{\ell N}|^2$ (left), or  $|U_{\ell N}|^4/\sum_{l^\prime} |U_{\ell^\prime N}|^2$ (right).
 } 
 \label{fig:width:tot}
\end{figure}

In Fig.~\ref{fig:width:tot} (left) we show the ``canonical'' decay rates for the modes $N \to \ell^+ n \pi$ and the inclusive estimate given by $N \to \bar u d \ell^+$  as a function of the neutrino mass $m_N$ (``canonical'' here means that all lepton mixing elements $|U_{\ell N}|$ are removed from the expressions). In Fig.~\ref{fig:width:tot} (right) we show the ``canonical'' branching ratios for the full processes $W^+\to \ell^+ N\to \ell^+ \ell^+ n\pi$ and $W^+\to \ell^+ N\to \ell^+ \ell^+ \bar u d$. The actual rates (left) and branching ratios (right) can be obtained by multiplying these canonical values by 
$|U_{\ell N}|^2$ (left) and $|U_{\ell N}|^4/\sum_{l^\prime} |U_{\ell^\prime N}|^2$ (right), respectively.  These factors can be easily deduced considering the expressions for the partial $N$ decay rates in Eqs.~\eqref{eq:width:1pi}, \eqref{eq:width} and \eqref{eq:width:3h}, and the expression for the neutrino 
width $\Gamma_N$ 
\cite{Atre:2009rg,Dib:2015oka,Arbelaez:2017zqq}:
\begin{equation}
\Gamma_{N}\simeq 1.1 \times \frac{G_{F}^{2}}{12\pi^{3}}m_{N}^{5} \sum_{\ell} |U_{ \ell N}|^{2} ,
\label{eq:GN1}
\end{equation}
which is a good approximation in the range  5 GeV$< m_N < 20$ GeV.

From Fig.\,\ref{fig:width:tot}, it can be seen that $\Gamma(N\to 3\pi \ell)/\Gamma(N \to 2 \pi \ell)\approx 0.7$ . This  is expected, as modes with more particles in the final state tend to be smaller than those with fewer particles. However, Fig.~\ref{fig:width:tot} also shows that the \textit{single} pion mode is suppressed with respect to the \hbox{two-pion} mode, $\Gamma(N\to \pi \ell)/\Gamma(N \to 2\pi \ell)\approx 0.4$, thus contradicting the previous argument, but, as explained in the Introduction, the probability of producing a single pion from a pair of energetic quarks is small and constitutes an even stronger suppression that the two- or three-pion final state.

We would like to remark that the quark level process $N \to \bar u d \ell^+$ can be considered quite confidently as the inclusive semileptonic process --what is usually called the quark-hadron duality~\cite{Shifman:2000jv}-- 
for $m_N > 5$ GeV, because the pion multiplicity can be quite large for these invariant masses, and the region is well above the resonance region. 
This is not the case in $b \to c$ transitions at invariant masses near 5 GeV, where the hadronic modes are very few. Indeed, the recent BaBar and Belle data indicate that a sizable duality violation occurs in the $\Br_{\rm exp} (B \to D \tau \nu) + \Br_{\rm exp} ( B \to D^* \tau \nu) > \Br_{\rm SM} (B \to X_c \tau \nu) $ \cite{Freytsis:2015qca}. 
Finally, let us estimate the expected number of $W\to \ell N \to \ell \ell n\pi$ events. According to 
Ref.~\cite{Aad:2016naf}, at the end of the LHC Run II one may expect a sample of ${\cal N}_W \sim 10^9$ $W$ decays. 
Considering the canonical branching ratios in Fig.\,\ref{fig:width:tot} (right), we can estimate the minimal values of the lepton mixing element factor that would generate 5 events or more, for a benchmark value of 
$m_N= 10$~GeV. At this mass, the figure gives a canonical branching ratio 
\hbox{${\cal B}(W^+\to \pi^0\pi^- \mu^+ \mu^+ )\sim 3 \times 10^{-4}$}. The single pion and three pion modes are, as shown in Fig.~\ref{fig:width:tot}, slightly smaller: 
${\cal B}(W^+\to \pi^- \mu^+ \mu^+) \approx 0.4\,  {\cal B}(W^+\to \pi^0 \pi^- \mu^+ \mu^+)$ and 
${\cal B}(W^+\to\pi^+\pi^- \pi^- \mu^+ \mu^+) \approx 0.7\,  {\cal B}(W^+\to \pi^0 \pi^- \mu^+ \mu^+)$.
Including all the modes, we have 

\begin{align}
& {\cal B}(W^+\to \pi^- \mu^+ \mu^+, \pi^0\pi^- \mu^+ \mu^+, \pi^+\pi^-\pi^-\mu^+ \mu^+,\pi^0\pi^0\pi^-\mu^+ \mu^+)\nonumber\\
& \hspace{36pt} \equiv {\cal B}(W^+\to n\pi \mu^+\mu^+)
 \approx 8\times 10^{-4}. 
\end{align}

In order to obtain more than 5 events, we must have:
\[
{\cal N}_W\times {\cal B}(W^+\to n \pi  \mu^+ \mu^+) \times \frac{|U_{\mu N}|^4}{\sum_{\ell} |U_{\ell N}|^2 } > 5 ,
\]
which implies $|U_{\mu N}|^2 \gtrsim 6.2\times 10^{-6}$, provided other mixing elements are smaller. If instead all mixing elements are comparable, then this lower bound increases by a factor 3, i.e.  \hbox{$|U_{e N}|^2, 
\, |U_{\mu N}|^2, \, |U_{\tau N}|^2 \gtrsim 1.9 \times 10^{-5}$}. These results are ideal,  disregarding the required cuts and backgrounds. 
However, these bounds can be made about one order of magnitude stronger if one adds all both charges and lepton flavors $W^\pm\to n\pi \ell^\pm \ell^{\prime \pm}$ $(n=1,2,3)$, i.e.
$|U_{e N}|^2,  \, |U_{\mu N}|^2, \, |U_{\tau N}|^2 \gtrsim 2 \times 10^{-6}$. On the other hand, if no signal is found, this limit on the mixings would become the new upper bound.
Current upper bounds for $|U_{\ell N}|^2$ in the range 5 GeV$< m_N < 50 $ GeV come from DELPHI~\cite{Abreu:1996pa}, which are  $|U_{e N}|^2 , \,  |U_{\mu N}|^2 ,\,  |U_{\tau N}|^2  \lesssim 2.1 \times 10^{-5}$. 
Consequently, using the modes $W^\pm \to n \pi \ell^\pm \ell^{\prime \pm}$, a sample of $10^9$ produced $W$ bosons may be able to improve the current upper bound on $|U_{\ell N}|^2$ by nearly an order of magnitude.  Moreover, these pionic modes seem to be amongst  the most promising to explore the sterile neutrino mass region \hbox{5 GeV$< m_N< 20$ GeV}.

A final remark concerns the Majorana vs. Dirac signals. While the DELPHI bounds just mentioned apply to either Majorana or Dirac $N$, our studied modes do make a difference. So far we have considered modes with equal sign dileptons. These modes only occur if $N$ is Majorana. The modes with opposite charge leptons, i.e. $e^+ e^-, \mu^+\mu^-, \mu^+ e^-$ etc. will also occur and with similar rates as those above, provided $N$ is Majorana. Although the opposite charge dilepton modes suffer from other backgrounds, they may help improve the statistics for a Majorana $N$. On the other hand, if $N$ were a Dirac neutrino, all the equal-sign dilepton modes we analyzed will be absent, but the opposite charge dileptons  will still be present and with the same rates as calculated above.

\section{Conclusions}\label{sec:conclusion}

In this work we have proposed a sterile neutrino search for masses $m_N$ in the range $5-20\mbox{ GeV}$, which is a region where neither rare meson decays ($B$, $D$, etc.) nor $pp\to \ell\ell jj, \ell\ell\ell \nu$ modes at the LHC  are sensitive to the presence of such neutrinos. Our search is based on the exclusive semileptonic processes $W \to n \pi \ell\ell$ ($n= 1,2, 3, \ldots$).
We explicitly studied the rates for $W^\pm\to\pi^\mp\ell^\pm\ell^\pm$,  $W^\pm\to\pi^0 \pi^\mp\ell^\pm\ell^\pm$ and 
$W^\pm\to\pi^\pm \pi^\mp\pi^\mp\ell^\pm\ell^\pm, \, \pi^0\pi^0\pi^\mp\ell^\pm\ell^\pm$. 

We find, as expected, that the single pion mode is suppressed with respect to the two-pion and three-pion modes, while the latter two rates are comparable. This suppression is explained as the low probability of forming a single pion from the two energetic light quarks produced in the decay of $N$ with mass of a few GeV or more, and expressed as the small fraction $f_\pi/m_N$. 

In the two-pion mode, the calculation involves an extrapolation of the pion form factor at large time-like $q^2$. For this purpose we use two models of form factors, one based on the vector meson dominance model (VDM) used by the BaBar collaboration and another based on light front holographic QCD (LFH). The two treatments show a remarkable agreement in the extrapolated region, which may be considered both an indication of the success of these two treatments as well as a confidence on our estimate.

In the three-pion mode, we use an extrapolation from tau lepton decays based on the so-called CLEO parametrization, which includes a series of resonances. This mode gives a value that is consistently lower than the two-pion mode by a factor of $\sim 0.7$ for all the $m_N$ mass range, which is also a sign of confidence on the model estimate.


Assuming $\sim 10^9$ W's produced by the end of the run II of the LHC, and in the best-case scenario of negligible background, our studied pionic modes could be able to put limits to $|U_{\ell N}|^2 \sim 2\times 10^{-6}$, one order of magnitude more stringent than the bounds coming from DELPHI. Moreover, the equal sign dilepton modes will provide signals or bounds for Majorana $N$, while the study of opposite sign dileptons will provide bounds for a Dirac $N$, provided the equal sign modes are not seen. 
As a final comment, while we did not study the backgrounds, which may turn to be very difficult to eliminate considering the pion identification, one should still keep in mind that for sterile neutrino masses below $\sim 20 $ GeV, the neutrino may live long enough to cause an observable displacement between the primary and secondary lepton, a feature that may greatly help reduce the backgrounds. 

After finishing this study we learned of two recent works \cite{Nemevsek:2018bbt, Cottin:2018hyf} that deal with the issue of displaced vertices in $N$ related processes.

\section*{Acknowledgements}

We are grateful to Sebastian Tapia and Edson Carquin for useful comments. X.Y. thanks Han Ying for useful discussions. C.S.K. was supported by the National Research Foundation of Korea (NRF)
grant funded by Korea government of the Ministry of Education, Science and
Technology (MEST) No. 2011-0017430, No. 2011-0020333 and No. 2016R1D1A1A02936965. This work was supported in part by FONDECYT (Chile) grants 1170171 and 3170906, and CONICYT (Chile) PIA/Basal FB0821 and Ring ACT1406.

\bibliographystyle{JHEP}
\bibliography{ntopibib}

\end{document}